\documentclass[aps,pra,superscriptaddress,showpacs,showkeys,floatfix]{revtex4-1}
\usepackage{graphicx,color}
\usepackage{epsfig}
\usepackage{graphicx}
\usepackage{mathrsfs}
\usepackage{color}
\usepackage{bm}
\usepackage{amsmath,amssymb,amsthm,amscd}
\usepackage{bbold}
\usepackage{mdwlist}

 \newcommand{\bra}[1]{\langle{#1} |}
 \newcommand{\ket}[1]{|{#1}\rangle  }
 
 \newcommand{\ketbra}[2]{\vert {#1} \rangle \langle{#2}\vert}

\newcommand{\Tr}{\operatorname{Tr}}

\begin{document}

\title{Witnessing quantum capacities of correlated channels}

\author{Chiara Macchiavello}
\affiliation{Quit group, Dipartimento di Fisica, 
Universit\`a di Pavia, via A. Bassi 6, 
 I-27100 Pavia, Italy}
\affiliation{Istituto Nazionale di Fisica Nucleare, Gruppo IV, via A. Bassi 6,
  I-27100 Pavia, Italy}

\author{Massimiliano F. Sacchi}
\affiliation{Istituto di Fotonica e Nanotecnologie - CNR, Piazza Leonardo
  da Vinci 32, I-20133, Milano, Italy}
\affiliation{Quit group, Dipartimento di Fisica, 
Universit\`a di Pavia, via A. Bassi 6, 
 I-27100 Pavia, Italy}
	
\date{\today}

\begin{abstract} 
We test a general method to detect lower bounds of the quantum channel 
capacity for two-qubit correlated channels. 
We consider in particular correlated dephasing, depolarising and amplitude
damping channels. We show that the method is easily implementable, it does not 
require {\em a priori} knowledge about the channels, and it is very efficient,
since it does not rely on full quantum process tomography.  
\end{abstract}

\maketitle

\section{Introduction}
The property of a quantum communication channel to convey quantum information 
is quantified in terms of the quantum capacity $Q$ \cite{lloyd,barnum,devetak,hay},   
which corresponds to the maximum 
number of qubits that can be reliably transmitted per channel use. 
In any realistic scenario noise is unavoidably present and the amount of 
information that can be transmitted is lower than in the ideal noiseless
case.  
It is therefore important to develop efficient means to establish whether
the channel can still be profitably employed for 
information transmission in the presence of noise, that 
may be completely unknown. 
\par A standard method to infer the effect of noise on a communication channel 
relies on quantum process tomography \cite{proc}, 
but this, however, is a 
demanding procedure in terms of the number of different measurement settings
needed, since it scales as $d^4$ for a finite $d$-dimensional quantum system. 
In Ref. \cite{ourl} a method was recently proposed to gain some information 
on the channel ability to transmit quantum information by employing a smaller 
number of measurements, that scales as $d^2$.  
A lower bound on the quantum channel capacity was derived 
and it was shown that it can be experimentally
accessed with a simple procedure. Such a procedure can be applied to any unknown
quantum communication channel. The efficiency of the method was tested
for many examples of single qubit channels, and for the generalised Pauli
channel in arbitrary finite dimension. 

In this paper we generalise this detection method to correlated qubit channels
and test its efficiency in this case.
Correlated qubit channels were originally studied in terms of classical 
information transmission and it was shown that for certain ranges of the
correlation strengths the use of entanglement allows one to enhance the 
amount of transmitted information along the channel 
\cite{kn:2002-macchiavello-palma-pra}.
Quantum memory (or correlated) channels then attracted growing attention, 
and interesting new features emerged by modeling of 
relevant physical examples, including
depolarizing channels~\cite{MMM},
Pauli channels~\cite{mpv04,daems,dc}, dephasing
channels~\cite{hamada,dbf,ps,gabriela,lidar}, 
amplitude damping channels \cite{vsdamp,vs2},
Gaussian channels~\cite{cerf},
lossy bosonic channels~\cite{mancini,lupo}, 
spin chains~\cite{spins}, collision models~\cite{collision}
and a micro-maser model~\cite{micromaser} (for a recent review on quantum 
channels with memory effects see Ref.~\cite{memo_review}).

\par The paper is organized as follows. In Sec. II we review the
method of bounding the quantum capacity by means of the Shannon
entropy pertaining to a vector of probabilities that can be inferred
by performing few measurements on the output of the channel and a reference 
system. 
In the subsequent sections we apply the method to two-qubit correlated 
channels, considering explicitly the memory dephasing channel (Sec. III), 
the memory depolarizing channels (Sec. IV), 
and the fully correlated damping channel (Sec. V). 
We summarise the results of the paper in Sec. VI. 
 
\section{Detection method}
Let us consider a generic quantum channel ${\cal E}$  
acting on a single system, 
and define ${\cal E}_N= {\cal E}^{\otimes N}$, 
where $N$ represents the number of 
channel uses. 
The quantum capacity $Q$ is defined 
as \cite{lloyd,barnum,devetak,hay}
\begin{eqnarray} Q=\lim _{N\to \infty}\frac
{Q_N}{N}\;,\label{qn} 
\end{eqnarray} 
where
$Q_N = \max
_{\rho } I_c (\rho , {\cal E}_N)$, 
and $I_c(\rho , {\cal E}_N)$ denotes the coherent information 
\cite{schumachernielsen}
\begin{eqnarray} I_c(\rho , {\cal E}_N) = S[{\cal E}_N (\rho )] - S_e
(\rho, {\cal E}_N)\;.\label{ic} \end{eqnarray} 
In Eq. (\ref{ic}),
$S(\rho )=-\Tr [\rho \log _2 \rho ]$ is the von Neumann entropy, and
$S_e (\rho, {\cal E})$ represents the entropy exchange \cite{schumacher}, i.e.
$S_e (\rho, {\cal E})= S[({\cal I}_R \otimes {\cal
E})(|\Psi _\rho \rangle \langle \Psi _\rho |)] $,
where $|\Psi _\rho \rangle $ is any purification of $\rho $ by means of a 
reference quantum system $R$, namely 
$\rho =\Tr _R [|\Psi _\rho \rangle \langle \Psi _\rho|]$.

\par In Ref. \cite{ourl} we derived a lower bound for the quantum capacity $Q$ that can be 
easily accessed without requiring full process tomography of the quantum
channel. We briefly review the derivation here in the following.  
For any complete set of orthogonal projectors $\{\Pi _i\}$,  
one has \cite{NC00} 
$S(\rho )\leq  S(\sum _i \Pi _i \rho \Pi _i)$. Then, for any orthonormal basis 
$\{ |\Phi _i \rangle \}$ for the tensor product of the reference and the 
system Hilbert spaces, one has the following bound to the entropy exchange
\begin{eqnarray}
S_e\left (\rho , {\cal E} \right )\leq H (\vec p)\;,  
\label{se-bound}
\end{eqnarray}
where $H(\vec p)\equiv -\sum _i p_i \log_2 p_i$ denotes the Shannon entropy for the vector of the 
probabilities $\{p_i\}$,  with 
\begin{eqnarray}
p_i = \Tr [({\cal I}_R \otimes {\cal
E})(|\Psi _\rho \rangle \langle \Psi _\rho |) |\Phi _i \rangle\langle\Phi _i|] 
\;.
\label{pimeas}
\end{eqnarray}
From Eq. (\ref{se-bound}) one obtains the following chain of bounds 
\begin{eqnarray}
Q \geq Q_1 \geq I_c(\rho , {\cal E}_1)\geq S\left [{\cal E} (\rho )\right ]-H(\vec p) 
\equiv Q_{DET}
\;,\label{qvec}
\end{eqnarray}
which holds for any $\rho$ and $\vec p$.   
A lower bound $Q_{DET}$ to the quantum capacity of an unknown channel
can then be detected by  preparing a
bipartite pure state $|\Psi _\rho \rangle $ and sending it through the
channel ${\cal I} _R\otimes {\cal E}$, where the unknown channel
${\cal E}$ acts on one of the two subsystems. Suitable
local observables on the joint output state are then measured in order
to estimate $S\left [{\cal E} (\rho )\right ]$ and $\vec p$, and to compute
$Q_{DET}$. Typically, for a fixed measurement setting, one can infer
different vectors of probabilities pertaining to different sets of
orthogonal projectors, as will be shown in the following. 
Moreover, one could also adopt an adaptive
detection scheme to improve the bound (\ref{qvec}) by varying the
input state $|\Psi _\rho \rangle $. Since no information is given {\em a priori} 
about the communication channel, typically we always choose a maximally entangled 
input state, so that the reduced input $\rho $ has maximum input entropy.

We will assume that only the local observables  
$O_i\otimes O_i$
on the system and reference are measured, 
where $\{O_i\}$ is a tomographically complete set on the system alone. 
Notice that the above measurements allow one to measure 
$\{O_i\}$  on the system alone by ignoring
the statistics of the measurement results on the reference. In this way, 
a complete tomography of the system output state can be performed, and 
therefore the term $S\left [{\cal E} (\rho )\right ]$ in Eq. (\ref{qvec})
can be estimated exactly. 
Our goal is to 
optimize the bound $Q_{DET}$ given these resources. 
This procedure requires $d^2-1$ measurement settings with respect to a
complete process tomography, where $d^4 -1$ observables have to be
measured:  
this choice greatly simplifies the experimental setup to detect the quantum capacity.

Let us now consider explicitly the case of qubits with $\{O_i\}=\{
\sigma _x ,\sigma _y,\sigma _z \}$. By denoting the Bell states as
\begin{eqnarray}
&&\!\!\!\!\!| \Phi ^\pm \rangle =\frac {1}{\sqrt 2}(|00 \rangle \pm |11 \rangle )\,, 
\ \  |\Psi ^\pm \rangle =\frac {1}{\sqrt 2}(|01 \rangle \pm |10 \rangle )\,, 
\label{phipsi}
\end{eqnarray}
it can be proven \cite{ourl} that the local measurement settings 
$\{\sigma _x \otimes \sigma _x,\sigma _y \otimes 
\sigma _y,\sigma _z \otimes \sigma _z \}$
allow one to estimate the vector $\vec p$ pertaining to 
the projectors onto the following inequivalent bases
\begin{eqnarray}
B_1= &&\{ a |\Phi ^+ \rangle + b |\Phi ^- \rangle  , 
-b |\Phi ^+ \rangle + a |\Phi ^- \rangle  ,  \nonumber \\& & \
c |\Psi ^+ \rangle + d |\Psi ^- \rangle  , -d  |\Psi ^+ \rangle + c |\Psi ^- \rangle   
\}\;,
\label{b1} 
\\ 
B_2= &&\{ a |\Phi ^+ \rangle + b |\Psi ^+ \rangle  , 
-b |\Phi ^+ \rangle + a |\Psi ^+ \rangle  , \nonumber \\& & 
c |\Phi ^- \rangle + d |\Psi ^- \rangle  , -d  |\Phi ^- \rangle + c |\Psi ^- \rangle  
\}
\;,\label{b2} 
\\ 
B_3= &&\{ a |\Phi ^+ \rangle + i b |\Psi ^- \rangle  , 
i b |\Phi ^+ \rangle + a |\Psi ^- \rangle  , \nonumber \\& & 
c |\Phi ^- \rangle + i d |\Psi ^+ \rangle  , i d  |\Phi ^- \rangle + c |\Psi ^+ \rangle   
\}
\;,\label{b3} 
\end{eqnarray}
with $a,b,c,d$ real and such that $a^2+b^2=c^2+d^2=1$.

The probability vector $\vec p$ for each choice of basis is evaluated according to Eq. 
(\ref{pimeas}). In order to obtain the tightest bound in (\ref{qvec}) given the 
fixed local measurements $\{\sigma _x \otimes \sigma _x,\sigma _y \otimes 
\sigma _y,\sigma _z \otimes \sigma _z \}$, the Shannon entropy 
$H(\vec p)$ will be then minimised as a function of the bases (\ref{b1}-\ref{b3}), 
by varying the coefficients $a,b,c,d$ over the three sets. 
In an experimental scenario, after collecting the outcomes of the measurements 
$\{\sigma _x \otimes \sigma _x,\sigma _y \otimes 
\sigma _y,\sigma _z \otimes \sigma _z \}$, this optimisation 
step corresponds to classical processing of the measurement outcomes.

The simplification of choosing a restricted set of measurements may 
generally come at a cost, since the 
evaluated Shannon entropy $H(\vec p)$ in Eq. (\ref{qvec}) may give a
poor bound to the quantum capacity.  Even for a unitary
transformation a simplified measurement setting could be inefficient to
provide a detectable bound.  For example, a detection scheme for
qubits for the unitary channels 
\begin{eqnarray} U=\frac 12 \left( I + i
\sum _{\alpha =x,y,z} \epsilon _\alpha  \sigma _\alpha \right )\;, \qquad \epsilon _\alpha =\pm
1\;, 
\end{eqnarray} 
with input $|\Phi ^+ \rangle $ and measurement on
any of the bases (\ref{phipsi}--\ref{b3}) gives always a uniform
probability vector, hence $H(\vec p)=2$. In these cases it is mandatory
to adopt an adaptive detection scheme: clearly, by varying the input
state to $(I_R \otimes U^\dag )|\Phi ^+ \rangle $ one obtains $H(\vec
p)=0$ from the Bell basis (\ref{phipsi}), thus recovering the result
$Q_{DET}=1$. A further possibility is to support our method with 
efficient estimation methods for unitaries \cite{Koch}.

We remember that the bound we are providing also gives detectable lower 
bounds to the private information \cite{devetak,ourl} and the 
entanglement-assisted classical capacity \cite{thapliyal,hol2,ourl}.

\section{Correlated dephasing channel}
We consider a dephasing quantum channel that maps two-qubit input states $\rho$ onto 
\begin{equation}
\mathcal{E}(\rho)=\sum_{i_1,i_2}
A_{i_1, i_2} \rho A^\dagger_{i_1, i_2},
\quad \quad i_k=0,1,
\label{dephmemory}
\end{equation}
where Kraus operators $A_{i_1, i_2}$ are defined in
terms of the Pauli operators $\sigma_0= I$ and
$\sigma _1 = \sigma_z$ as follows 
\begin{equation}
A_{i_1, i_2}=\sqrt{p_{i_1,i_2}}B_{i_1,i_2},\quad
B_{i_1,i_2}\equiv\sigma_{i_1}^{(1)}\otimes
\sigma_{i_2}^{(2)},
\label{Krausnuses}
\end{equation}
with $\sum_{\{i_k\}} p_{i_1,i_2}=1$, and 
$\sigma_{i_k}^{(k)}$ acting on the $k$-th qubit. 

We describe the joint probabilities in Eq. (\ref{Krausnuses}) by a Markov 
chain~\cite{kn:2002-macchiavello-palma-pra,hamada}, namely  
\begin{equation}
p_{i_1,i_2}=p_{i_1}p_{i_2|i_1},
\label{eq:markov}
\end{equation}
with 
\begin{equation}
p_{i_2|i_1}=(1-\mu)\,p_{i_2}+\mu\,\delta_{i_1,i_2}.
\label{eq:propagator}
\end{equation}
The parameter $\mu\in[0,1]$ measures degree of correlation of the channel: 
it is the  probability that the same operator 
(either $I$ or $\sigma_z$) 
is applied for two consecutive uses
of the channel, whereas $1-\mu$ is the probability that 
the two operators are uncorrelated.
The limiting cases $\mu=0$ and 
$\mu=1$ 
correspond to memoryless channels and channels with 
perfect memory, respectively. 
The correlated 
dephasing channel is easily shown to be degradable \cite{ds}, hence $Q=Q_1$,  
and its quantum capacity is given by \cite{dbf,ps,physc}
\begin{eqnarray}
Q= \left \{2 - 
p\, H_2[(1-p)(1- \mu )]- (1-p)\, H_2 [p (1-\mu )]-  H_2 (p) \right \}
\;,\label{qdetmem}
\end{eqnarray}
where $p\equiv p_1$, and $H_2(p)\equiv -p\log_2 p -(1-p)\log_2 (1-p)$ 
denotes the binary Shannon entropy.  Notice also that Eq. (\ref{qdetmem}) is invariant 
by replacing $p$ with $(1-p)$.

\begin{figure}[htb]
  \includegraphics{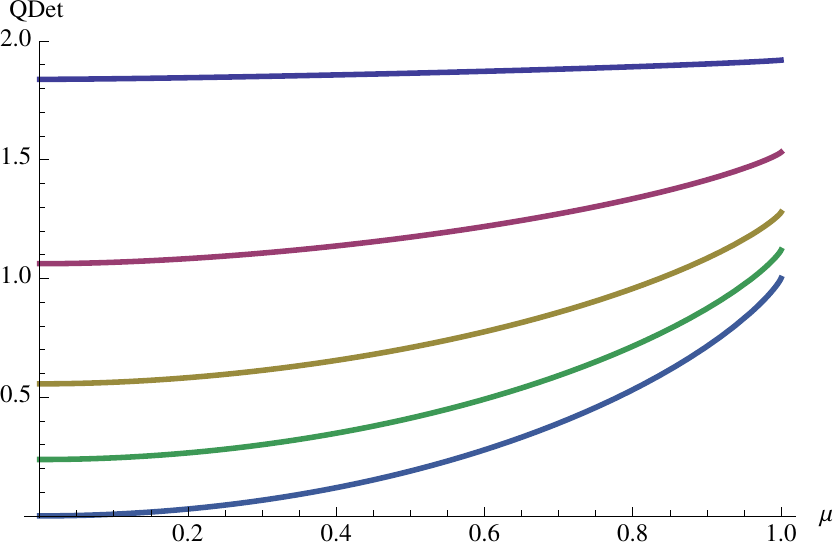}
  \caption{Detected quantum capacity for the correlated dephasing channel 
versus the correlation parameter $\mu $ for different values of the probability 
$p$ (from top to bottom $p=0.01,\ 0.1,\ 0.2,\ 0.3,\ 0.5$).  
Two maximally entangled input states are used and Bell measurements are 
considered. The curves coincide with the quantum capacity given by Eq. 
(\ref{qdetmem}).}
  \label{memodeph}
\end{figure}

We consider now a detection scheme with two input qubits $A$ and $B$  which are maximally 
entangled with two reference qubits $R_A$ and 
$R_B$, namely a global input state $\ket{\Phi ^+}_{R_A,A} \ket{\Phi ^+}_{R_B,B}$. 
The corresponding output state is given by 
\begin{eqnarray}
\Xi &=&{\cal I}_{R_A}\otimes {\cal I}_{R_B}
\otimes {\cal E}_2 (\ket{\Phi ^+} \bra{\Phi ^+}_{R_A,A} \otimes 
\ket{\Phi ^+} \bra {\Phi ^+}_{R_B,B}) \nonumber \\&= &
(1-p)[(1-p)(1-\mu)+\mu ] 
\ket{\Phi ^+} \bra{\Phi ^+}_{R_A,A} \otimes 
\ket{\Phi ^+} \bra {\Phi ^+}_{R_B,B} \nonumber \\&+ &
p[p(1-\mu)+\mu ] 
\ket{\Phi ^-} \bra{\Phi ^-}_{R_A,A} \otimes 
\ket{\Phi ^-} \bra {\Phi ^-}_{R_B,B} \nonumber \\&+&
p(1-p)(1-\mu) 
\ket{\Phi ^+} \bra{\Phi ^+}_{R_A,A} \otimes 
\ket{\Phi ^-} \bra {\Phi ^-}_{R_B,B} \nonumber \\&+ &
p(1-p)(1-\mu) 
\ket{\Phi ^-} \bra{\Phi ^-}_{R_A,A} \otimes 
\ket{\Phi ^+} \bra {\Phi ^+}_{R_B,B} 
\;.\label{rhofulldeph}
\end{eqnarray} 
The reduced input state for qubits $A$ and $B$ 
is simply $\frac 14 I_A \otimes I_B$, and it remains invariant under 
the action of ${\cal E}$, hence 
the reduced output entropy equals 2 bits.   
We consider a measurement scheme on the output state (\ref{rhofulldeph}) 
where the set of observables $ \sigma _x 
\otimes \sigma _x $, $ \sigma _y \otimes \sigma _y $, and $ \sigma _z 
\otimes \sigma _z $ are measured on both couples 
of qubits $R_A,A$ and $R_B,B$. Such a scheme 
provides the vector of probabilities 
\begin{eqnarray}
\vec p= \{ (1-p)[(1-p)(1-\mu)+\mu ] , p[p(1-\mu)+\mu ] ,p(1-p)(1-\mu) ,p(1-p)(1-\mu) \}\;.
\label{pmuu}
\end{eqnarray}
A straightforward calculation shows that the detected quantum capacity coincides 
with the quantum capacity, namely
\begin{eqnarray}
Q\equiv Q_{DET}=2 -H (\vec p)\;.
\end{eqnarray}
Our detected bound provides exactly the quantum capacity, 
since $Q=Q_1$ due to the degradability of the channel, 
and the components of the vector $\vec p$ in Eq. (\ref{pmuu}) 
correspond to the eigenvalues 
of the joint output state (\ref{rhofulldeph}).  
In Fig. (\ref{memodeph}) we plot the detected capacity (\ref{qdetmem}) 
versus the correlation parameter $\mu $, for the following values 
$p=0.01,\ 0.1,\ 0.2,\ 0.3,\ 0.5$ 
(or, equivalently, $p=0.99,\ 0.9,\ 0.8,\ 0.7,\ 0.5$). 

\section{Correlated depolarizing channel} 

We study  the following correlated
depolarizing quantum channel \cite{kn:2002-macchiavello-palma-pra} 
that maps two-qubit input states $\rho$ onto 
\begin{equation} \mathcal{E}(\rho)=\sum_{i_1,i_2}
A_{i_1,i_2} \rho A^\dagger_{i_1,i_2}, \quad \quad i_k=0,1,2,3
\label{depomemory} \end{equation} 
where Kraus operators are defined as in Eq. (\ref{Krausnuses}), now with  
$\sigma_0=
I\,,\ \sigma _1 = \sigma_z \,, \ \sigma _2 = \sigma _x\,,\ \sigma _3 =
\sigma _y$.  The joint probabilities still satisfy the Markov chain rule as in
Eqs. (\ref{eq:markov},\ref{eq:propagator}), with $p_0=1-p$ and $p_1=p_2=p_3=\frac p3 $.

As in the previous case, the parameter $\mu\in[0,1]$ measures the degree of  
correlation of the channel: it is the probability that 
the same operator $\sigma _i $ is applied for two consecutive uses
of the channel, whereas $1-\mu$ is the probability that 
the two operators are uncorrelated.
Again, the limiting cases $\mu=0$ and 
$\mu=1$ 
correspond to memoryless channels and channels with 
perfect correlation, respectively. 

Let us consider now two input qubits $A$ and $B$  which are maximally 
entangled with two reference qubits $R_A$ and 
$R_B$, namely an input $\ket{\Phi ^+}_{R_A,A} \ket{\Phi ^+}_{R_B,B}$. 
We also rename the Bell states as follows 
\begin{eqnarray}
&&
\ket{\Phi _0}\equiv \ket{\Phi ^+}\;,
\nonumber \\& & 
\ket{\Phi _1}\equiv \ket{\Phi ^-}\;,\nonumber \\& & 
\ket{\Phi _2}\equiv \ket{\Psi ^+}\;,\nonumber \\& & 
\ket{\Phi _3}\equiv \ket{\Psi ^-}
\;.
\end{eqnarray}
The output state can then be written as 
\begin{eqnarray}
\Xi &=&{\cal I}_{R_A}\otimes {\cal I}_{R_B}
\otimes {\cal E} (\ket{\Phi _0} \bra{\Phi _0}_{R_A,A} \otimes 
\ket{\Phi _0} \bra {\Phi _0}_{R_B,B}) \nonumber \\&= &
\sum_{i,j=0}^3 p_{ij}\ket{\Phi _i} \bra{\Phi _i}_{R_A,A} \otimes 
\ket{\Phi _j} \bra {\Phi _j}_{R_B,B} 
\;,\label{outxi}
\end{eqnarray}
where
\begin{eqnarray}
&&p_{00}=(1-\mu) (1-p)^2 
+ \mu (1-p) \,,\nonumber \\& & 
p_{ii}=(1-\mu) \left ( \frac{p}{3}\right )^2 
+\mu  \frac p3 \,, \qquad i=1,2,3 \,,\nonumber \\& & 
p_{ij}= (1-\mu) \left ( \frac{p}{3}\right )^2\,, \qquad i,j=1,2,3 \quad i\neq j \,,\nonumber \\& & 
p_{0i}=p_{i0}=(1-\mu) \frac p3 (1-p)\,, \qquad i=1,2,3
\;.\label{pij16}
\end{eqnarray}
The reduced input state is simply $\frac 14 I_A \otimes I_B$, 
and remains invariant under the action of ${\cal E}$, hence 
the reduced output entropy equals 2 bits.   
A measurement scheme on the output state (\ref{outxi}) 
where the set of observables $ \sigma _x 
\otimes \sigma _x $, $ \sigma _y \otimes \sigma _y $, and $ \sigma _z 
\otimes \sigma _z $ are measured on both couples of qubits 
$R_A,A$ and $R_B,B$ 
provides all probabilities $p_{ij}$ in Eq. (\ref{pij16}). 

Then, we can write our detected bound as follows 
\begin{eqnarray}
Q\geq Q_{DET}= [2 -H(\{p_{ij} \})]= 
2 + p_{00}\log_2 p_{00} + 3 p_{11} \log_2 p_{11} + 6 p_{12}\log_2 p_{12}
+ 6 p_{01}\log_2 p_{01}
\;.\label{qdetmem2}
\end{eqnarray}

The detected capacity $Q_{DET}$ coincides with the maximum of the coherent
information evaluated in Ref.  \cite{physc} for a single use of the
memory channel (\ref{depomemory}). Since the channel is not
degradable, $Q_{DET}$ is just a lower bound of the quantum channel
capacity, whose exact expression is still unknown.

We notice, however, that for the fully correlated channel, i.e. $\mu =1$,
Kraus operators $\{\sigma _i \otimes \sigma _i \}$ are a commuting set, 
hence the channel is degradable \cite{ds} and one has 
\begin{eqnarray} 
Q\equiv Q_{DET}=Q=2-H_2(p)- p \log _2 3 \;,\label{fuldep}
\end{eqnarray}
which corresponds to the exact quantum capacity, that is therefore 
efficiently detected by our method. The result of Eq. (\ref{fuldep}) can be easily generalized 
to the case of fully correlated depolarized channels for qudits, thus giving 
\begin{eqnarray} 
Q\equiv Q_{DET}=Q=d-H_2(p)- p \log _2 (d^2 -1) \;.\label{fuldep2}
\end{eqnarray}

\begin{figure}[htb]
  \includegraphics{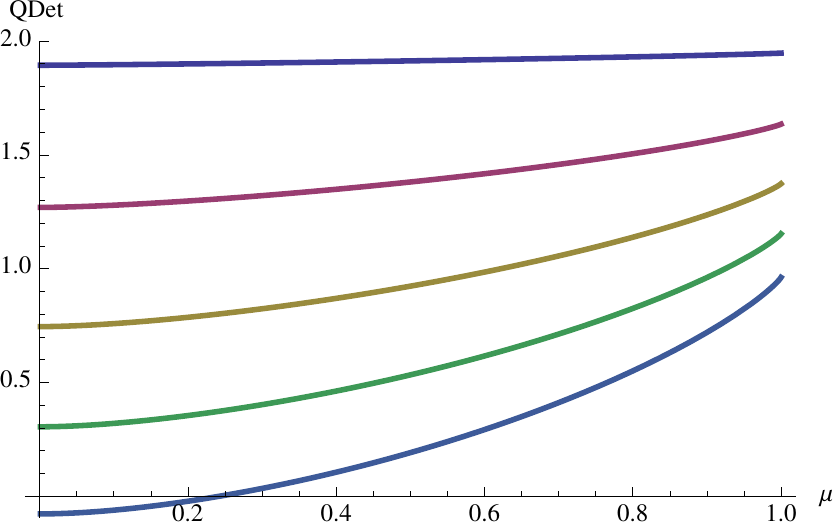}
  \caption{Detected quantum capacity for the correlated depolarizing 
channel versus memory parameter $\mu $ for different values of the probability 
$p$ (from top to bottom $p=0.005,\ 0.05,\ 0.1,\ 0.15,\ 0.2$). 
The detected quantum capacity is given by Eq. (\ref{qdetmem2})  
using two maximally entangled input states and Bell measurement. }
  \label{memodepo}
\end{figure}

In Fig. (\ref{memodepo}) we plot the detected bound (\ref{qdetmem2}) 
versus the correlation parameter $\mu $, for the following values 
$p=0.005,\ 0.05,\ 0.1,\ 0.15,\ 0.2$. 

\section{Fully correlated damping channel}

In this section we consider the following correlated amplitude damping channel 
acting on two qubits \cite{vsdamp}
\begin{equation}
{\cal E} (\rho)=\sum_{i=1}^2  B_i \,\rho\,B_i^\dag,
 \label{eq:full-memory-channel}
\end{equation}
with Kraus operators 
\begin{equation}
{B}_1 = \left(
\begin{array}{cccc}
  1 & 0 & 0 & 0  \\
  0 & 1 & 0 & 0  \\
  0 & 0 & 1 & 0  \\
  0 & 0 & 0 & \sqrt{\eta}  \\
  \end{array} \right), \,\,
\qquad {B}_2 = \left(
\begin{array}{cccc}
  0 & 0 & 0 & \sqrt{1-\eta}  \\
  0 & 0 & 0 & 0  \\
  0 & 0 & 0 & 0  \\
  0 & 0 & 0 & 0  \\
  \end{array} \right),
\label{eq:memory-Kraus-Operators}
\end{equation}
where the ordered basis $\{ \ket{00},   \ket{01},   \ket{10},   \ket{11} \}$ has been used.  
This channel describes a fully correlated damping, namely 
only the state $\ket{11}$ undergoes decay to $\ket{00}$ with probability 
$\eta $, 
while the other states $\ket{00},   \ket{01},   \ket{10}$ remain unaltered. 

We only consider the fully correlated case, because for 
partially correlated amplitude damping channels just numerical bounds on 
the quantum capacity are known \cite{vs2}. On the other hand, the fully 
correlated amplitude damping channel 
has been shown to be degradable for $\eta \geq 1/2$, and its quantum capacity is 
explicitly obtained by the following maximization \cite{vsdamp}
\begin{eqnarray}
  Q&=& \max_{\alpha,\beta,\delta}\Big\{
   -[\alpha+(1-\eta)\delta]\log_2[\alpha+(1-\eta)\delta]
- 2\beta\log_2\beta\,-\,\eta \delta \log_2\eta\delta\, +\,\nonumber\\
&&    + [1-(1-\eta)\delta]\log_2[1-(1-\eta)\delta]
+\,(1-\eta)\delta\log_2[(1-\eta)\delta]
\Big\},
\label{eq:QuantumCapacity-1}
\end{eqnarray}
with the constraints $\alpha+2\beta+\delta=1$ and $\alpha ,\beta ,\delta \geq 0$. 
For $\eta \leq 1/2$, one simply has $Q=\log _2 3$, corresponding just to coding 
on the noiseless subspace spanned by $\{ \ket{00},   \ket{01},   \ket{10} \}$. 

As in the previous examples, we 
consider an input maximally entangled state between the two
qubits $A$ and $B$ with two reference qubits $R_A$ and $R_B$, 
namely $\ket{\Phi ^+}_{R_A,A} \ket{\Phi ^+}_{R_B,B}$.
The output state is then given by
\begin{eqnarray}
\Xi = {\cal I}_{R_A}\otimes {\cal I}_{R_B}\otimes {\cal E}(\ket{\Phi ^+} \bra{\Phi ^+}_{R_A,A} \otimes 
\ket{\Phi ^+} \bra {\Phi ^+}_{R_B,B})
\;.\label{rhofull}
\end{eqnarray} 

Notice that Kraus operators $B_1$ and $B_2$ in Eq. (\ref{eq:memory-Kraus-Operators}) 
can be rewritten as 
\begin{eqnarray}
&&B_1 = \frac {3+\sqrt{\eta }}{4} I_A \otimes I_B +
\frac {1-\sqrt{\eta }}{4} \sigma _{zA} \otimes I_B +
\frac {1-\sqrt{\eta }}{4} I_A \otimes \sigma _{zB} 
-\frac {1-\sqrt{\eta }}{4} \sigma _{zA} \otimes \sigma _{zB} 
\;\nonumber \\& & 
B_2 = \frac {\sqrt{1-\eta }}{4} 
(\sigma _{xA}+ i \sigma _{yA})\otimes (\sigma _{xB}+ i \sigma _{yB}) \;.
\end{eqnarray}

It follows that the output state (\ref{rhofull}) has a block-diagonal form, 
i.e. $\Xi = \Xi _1 \oplus \Xi _2$, 
with 
\begin{eqnarray}
\Xi _1 &=&  B_1 (\ket{\Phi ^+} \bra{\Phi ^+}_{R_A,A} \otimes 
\ket{\Phi ^+} \bra {\Phi ^+}_{R_B,B})
B_1 ^\dag  \nonumber \\&= &   
y_\eta 
\left(
\begin{array}{cccc}
  x_\eta ^2 & x_\eta  & x_\eta  & x_\eta   \\
  x_\eta  & 1 & 1 & 1  \\
  x_\eta  & 1 & 1 & 1  \\
  x_\eta  & 1 & 1 & 1  \\
  \end{array} \right)
\;,\label{1block}
\end{eqnarray} 
on the ordered basis $\{ \ket{\Phi ^+}_{R_A,A} \ket{\Phi ^+}_{R_B,B}, 
\ket{\Phi ^+}_{R_A,A} \ket{\Phi ^-}_{R_B,B}, \ket{\Phi ^-}_{R_A,A} \ket{\Phi ^+}_{R_B,B}, 
\ket{\Phi ^-}_{R_A,A} \ket{\Phi ^-}_{R_B,B}\} $, with
\begin{eqnarray}
y_\eta =\frac {(1-\sqrt {\eta })^2}{16}, \qquad x_\eta =\frac{3+\sqrt {\eta }}{1-\sqrt{\eta}}
\;,
\end{eqnarray}
whereas 
\begin{eqnarray}
\Xi _2 = B_2 
(\ket{\Phi ^+} \bra{\Phi ^+}_ {R_A,A} \otimes 
\ket{\Phi ^+} \bra {\Phi ^+}_{R_B,B}) 
B_2 ^\dag = \frac {1-\eta }{4}
{\ketbra {10}{10}}_{R_A,A}\otimes 
{\ketbra {10}{10}}_{R_B,B}
\;.
\end{eqnarray}

The reduced output state is given by 
\begin{eqnarray}
{\cal E}\left ( \frac {I_A}{ 2} \otimes \frac {I_B} {2} \right )=
\frac 14 \left(
\begin{array}{cccc}
  2-\eta  & 0 & 0 & 0  \\
  0 & 1 & 0 & 0  \\
  0 & 0 & 1 & 0  \\
  0 & 0 & 0 & \eta  \\
  \end{array} \right)\,,
\label{outeta}
\end{eqnarray}
on the ordered basis $\{ \ket{00},   \ket{01},   \ket{10},   \ket{11} \}$. 
Notice that the present channel is clearly an example of non-unital channel. 

We consider now a detection scheme on the output state 
where the set of observables $ \sigma _x 
\otimes \sigma _x $, $ \sigma _y \otimes \sigma _y $, and $ \sigma _z 
\otimes \sigma _z $ are measured on both couples of qubits $R_A,A$ and $R_B,B$. 
The set of probabilities that can be obtained by this measurement setting and 
minimizes the Shannon entropy $H(\vec p)$, 
corresponds to the set of projectors on the following 
states
\begin{eqnarray}
&&{\ket{00}}_{R_A,A}{\ket{01}}_{R_B,B}\;,
{\ket{00}}_{R_A,A}{\ket{10}}_{R_B,B}\;,
{\ket{11}}_{R_A,A}{\ket{01}}_{R_B,B}\;,
{\ket{11}}_{R_A,A}{\ket{10}}_{R_B,B}\;,\nonumber \\& & 
{\ket{01}}_{R_A,A}{\ket{00}}_{R_B,B}\;,
{\ket{01}}_{R_A,A}{\ket{01}}_{R_B,B}\;,
{\ket{01}}_{R_A,A}{\ket{10}}_{R_B,B}\;,
{\ket{01}}_{R_A,A}{\ket{11}}_{R_B,B}\;,\nonumber \\& & 
{\ket{10}}_{R_A,A}{\ket{00}}_{R_B,B}\;,
{\ket{10}}_{R_A,A}{\ket{01}}_{R_B,B}\;,
{\ket{10}}_{R_A,A}{\ket{11}}_{R_B,B}
\;,\label{11}
\end{eqnarray}
for which $p=0$, 
\begin{eqnarray}
{\ket{10}}_{R_A,A}{\ket{10}}_{R_B,B}
\;,\label{1010}
\end{eqnarray}
for which $p=\frac {1-\eta}{4}$, 
and \cite{nota}
\begin{eqnarray}
&&
|\chi _1\rangle 
\equiv (a {\ket {\Phi^+}}_{R_A,A} + b {\ket {\Phi^-}}_{R_A,A})
(a {\ket {\Phi^+}}_{R_B,B} + b {\ket {\Phi^-}}_{R_B,B})\;,
\nonumber \\& & 
|\chi _2 \rangle \equiv 
(a {\ket {\Phi^+}}_{R_A,A} + b {\ket {\Phi^-}}_{R_A,A})
(-b  {\ket {\Phi^+}}_{R_B,B} + a {\ket {\Phi^-}}_{R_B,B})\;,
\nonumber \\& & 
|\chi _3 \rangle \equiv 
(-b {\ket {\Phi^+}}_{R_A,A} + a {\ket {\Phi^-}}_{R_A,A})(a {\ket {\Phi^+}}_{R_B,B} + b {\ket {\Phi^-}}_{R_B,B})
\;,\nonumber \\& & 
|\chi _4 \rangle \equiv (-b  {\ket {\Phi^+}}_{R_A,A} + a {\ket {\Phi^-}}_{R_A,A})
(-b  {\ket {\Phi^+}}_{R_B,B} + a {\ket {\Phi^-}}_{R_B,B})\;,
\label{44}
\end{eqnarray}
with $a$ and $b$ real, such that $a^2+b^2=1$ and $-\sum _{i=1} ^4 
q_i \log _2 q_i$ is minimized, where 
\begin{eqnarray}
q_i = \langle \chi _i | \Xi_1 | \chi _i \rangle 
\;.
\end{eqnarray}

We can now write the detection bound as follows 
\begin{eqnarray}
Q\geq Q_{DET}= H({\vec s }\,) -  H_2\left (\frac{1-\eta }{4}\right ) - 
H({\vec q }\,)\;,\label{qdetfull}
\end{eqnarray}
where $\vec s =\{ (2-\eta )/4, 1/4, 1/4, \eta /4 \}$ corresponds to the eigenvalues of the output reduced state (\ref{outeta}), and $\vec q =\{q_1,q_2,q_3,q_4 \}$.  

\begin{figure}[htb]
  \includegraphics{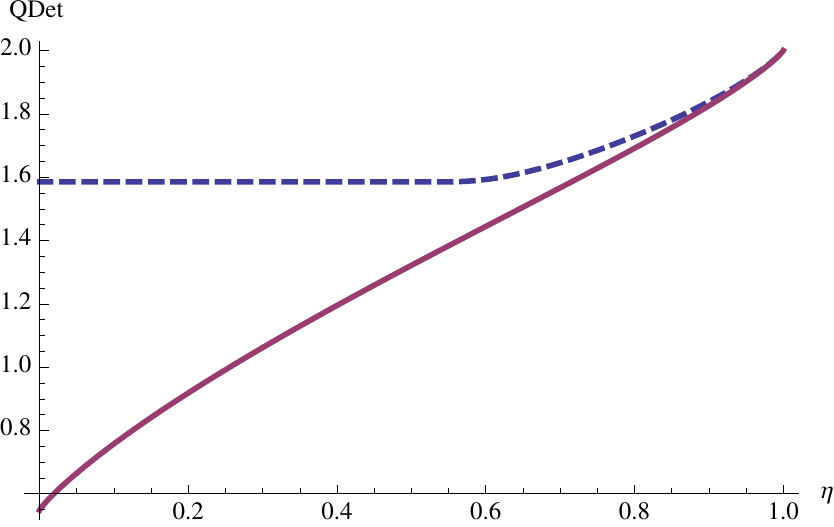}
  \caption{Fully correlated amplitude damping channel with parameter $\eta $: 
detected quantum capacity (thick line) 
with maximally entangled input and projective measurement on states 
(\ref{11},\ref{1010},\ref{44}), along with the theoretical 
quantum capacity (dashed line) given by Eq. (\ref{eq:QuantumCapacity-1}).}
  \label{figfull}
\end{figure}

In Fig. \ref{figfull} we plot the detection bound along with the quantum 
capacity of the fully correlated amplitude
damping channel versus damping parameter $\eta $.  The looseness of the bound for $\eta < 1/2$ is 
due to the fact the input maximally entangled state is very suboptimal for strong damping. Notice, however, that 
the positivity of the quantum  capacity is witnessed for all values of $\eta $.

\section{Conclusions} 
We have applied  a general method to witness lower 
bounds to the quantum capacity of quantum communication channels 
developed in Ref. \cite{ourl}  to the case of correlated qubit channels. 
We have shown that our  method
does not require any a priori knowledge about the channel itself and relies
on a number of measurement settings that scales more favorably
with respect to full process tomography.  Specifically, we tested
the method on two-qubit correlated channels of dephasing,
depolarizing and amplitude damping type, and showed that a fixed
maximally entangled input state of two system qubits and two reference qubits, and 
a setting of local measurements allow one to certify the quantum capacity, 
without the need of a complete tomographical reconstruction of the channel 
operation.
We want to emphasize that for quantum optical systems our method is easily 
implementable with present-day technologies \cite{tech}.


\begin{thebibliography}{99}
\bibitem{nota} For the subspace spanned by $\{ \ket{\Phi ^+}_{R_A,A} \ket{\Phi ^+}_{R_B,B}, 
\ket{\Phi ^+}_{R_A,A} \ket{\Phi ^-}_{R_B,B}, \ket{\Phi ^-}_{R_A,A} \ket{\Phi ^+}_{R_B,B}, 
\ket{\Phi ^-}_{R_A,A} \ket{\Phi ^-}_{R_B,B}\} $, 
the detection scheme where the set of observables $ \sigma _x 
\otimes \sigma _x $, $ \sigma _y \otimes \sigma _y $, and $ \sigma _z 
\otimes \sigma _z $ are measured on both couples of qubits $R_A,A$ and $R_B,B$ allows one to 
generally estimate the probabilities pertaining to projectors on states of the form
\begin{eqnarray}
|\psi  \rangle 
= A  \ket{\Phi ^+}_{R_A,A} \ket{\Phi ^+}_{R_B,B} +B 
\ket{\Phi ^+}_{R_A,A} \ket{\Phi ^-}_{R_B,B} +C \ket{\Phi ^-}_{R_A,A} \ket{\Phi ^+}_{R_B,B} 
+ D \ket{\Phi ^-}_{R_A,A} \ket{\Phi ^-}_{R_B,B}
\;,\nonumber
\end{eqnarray}
with $A,B,C,D$ real, and satisfying $A^2+B^2+C^2+D^2=1$ along with $AD=BC$. 
This is equivalent to factorized states in the $A-B$ cut, namely 
\begin{eqnarray}
|\psi  \rangle   
\equiv (a {\ket {\Phi^+}}_{R_A,A} + b {\ket {\Phi^-}}_{R_A,A})\otimes 
(c {\ket {\Phi^+}}_{R_B,B} + d {\ket {\Phi^-}}_{R_B,B})\;,
\nonumber
\end{eqnarray}

with $a=A/\sqrt{A^2 +C^2}\,, \ b=C/\sqrt{A^2 +C^2}\,,\ c=\sqrt{A^2 +C^2}\,,$ and 
$d=B\sqrt{A^2 +C^2}/A$. 
 
Moreover, the symmetry of Eq. (\ref{1block}), i.e. the invariance for exchange of systems $A$ 
with $B$, allows us to optimize just over the set of projectors in Eq. (\ref{44}).
\end{thebibliography}
\end{document}